\documentclass[10pt,conference]{IEEEtran}
\IEEEoverridecommandlockouts
\usepackage{cite}
\usepackage{amsmath,amssymb,amsfonts}
\usepackage{algorithmic}
\usepackage{graphicx}
\usepackage{textcomp}
\usepackage{xcolor}
\def\BibTeX{{\rm B\kern-.05em{\sc i\kern-.025em b}\kern-.08em
    T\kern-.1667em\lower.7ex\hbox{E}\kern-.125emX}}
    
\usepackage{pifont}
\usepackage{multirow}
\usepackage{flushend}
\usepackage{caption}
\usepackage{booktabs}
\usepackage{url}

\begin{document}

\title{Towards Benchmarking GUI Compatibility Testing on Mobile Applications
}

\author{
\IEEEauthorblockN{Jiaming Ye\IEEEauthorrefmark{1},
Mulong Xie\IEEEauthorrefmark{2},
Siyuan Chen\IEEEauthorrefmark{1},
Fuyuan Zhang\IEEEauthorrefmark{1},
Lei Ma\IEEEauthorrefmark{3},
Zhenchang Xing\IEEEauthorrefmark{4} and,
Jianjun Zhao\IEEEauthorrefmark{1}
}
\IEEEauthorblockA{
\IEEEauthorrefmark{1}Kyushu University, Japan\\
Email: ye.jiaming.852@s.kyushu-u.ac.jp}
\IEEEauthorblockA{
\IEEEauthorrefmark{2}Australian National University, Australia\\}
\IEEEauthorblockA{
\IEEEauthorrefmark{3}University of Alberta, Canada\\}
\IEEEauthorblockA{
\IEEEauthorrefmark{4}Data61, CSIRO, Australia\\
}
}

\maketitle

\begin{abstract}

GUI is a bridge connecting user and application. Existing GUI testing tasks can be categorized into two groups: functionality testing and compatibility testing. While the functionality testing focuses on detecting application runtime bugs, the compatibility testing aims at detecting bugs resulting from device or platform difference.
To automate testing procedures and improve testing efficiency, previous works have proposed dozens of tools. To evaluate these tools, in functionality testing, researchers have published testing benchmarks. Comparatively, in compatibility testing, the question of ``Do existing methods indeed effectively assist test cases replay?'' is not well answered. To answer this question and advance the related research in GUI compatibility testing, we propose a benchmark of GUI compatibility testing. In our experiments, we compare the replay success rate of existing tools. Based on the experimental results, we summarize causes which may lead to ineffectiveness in test case replay and propose opportunities for improving the state-of-the-art.

\end{abstract}
\section{Introduction}

Android, as well as Android devices have taken the first place of global mobile market, with over 71\% share of it, and leave its competitors (e.g., the iOS devices) far behind~\cite{android-market-share}. The rapid development of mobile hardware (e.g., mobile chips, cameras) has empowered Android developers to create exciting applications. Nowadays, the Android applications are continuously changing people's daily life.
To ensure the application reliability and user experience, the Android applications must follow the software lifecycle like classic C/C++ programs. That is, the Android applications need to be thoroughly tested to ensure a good user experience. Among all application testing tasks, GUI testing is one of the most important tasks.


Existing GUI testing tasks can be categorized into two groups: functionality testing and compatibility testing. The functionality testing focuses on detecting bugs on applications (e.g., crashes, unexpected quits)~\cite{glib}. To improve the testing efficiency, researchers propose approaches to automate GUI testing. In functionality testing, Su \emph{et al.}~\cite{stoat} and Mao \emph{et al.}~\cite{sapienz} propose model-based and search-based application testing approaches. Their methods try to traverse as much GUIs as possible, so that the bugs can be thoroughly exposed. Based on these methods and the efforts of developers, existing testing tools are evaluated in related benchmarks~\cite{themis} and widely adopted in industries.

Generally, compatibility testing aims at detecting bugs resulting from device or platform difference, and further keeping a uniform user experience. Current compatibility testing methods can be categorized into intrusive testing and non-intrusive testing. Specifically, the intrusive testing relies on extracting GUI information (e.g., widget coordinates, paths) by ADB (i.e., Android Developer Bridge) while the non-intrusive testing endeavors to reduce the participation of ADB. As the ``One time development, multi-terminal deployment'' becomes a trend among developers (e.g., Harmony OS)~\cite{harmonyos}. The compatibility testing attracts more and more attentions from industries and academics.

Compared with functionality testing, there is a gap between proposed method and practice. This gap is not thoroughly exposed in research experiments but can not be ignored in industrial testing tasks. Particularly, though previous researchers have proposed a number of tools~\cite{themis}  to automate testing, these tools are not widely adopted in industries~\cite{jiamingyepaper}. Existing industrial compatibility testing tasks are mainly done by intensive manpower. The developers manually repeat test cases on devices, which leads to overwhelming efforts to update test cases on new devices because the hardware differences between tested devices. ``Do existing methods indeed effectively assist test cases replay?'', such question remains unanswered. The only way to answer this question is conducting comprehensive evaluation on a benchmark. However, considering the lack of benchmark for compatibility testing, building a benchmark for evaluating tools is in urgent needs. 



In this paper, we focus on three points towards building a benchmark for compatibility testing: dataset with test cases, evaluations and discussions. We first build a dataset including 30 test cases from 8 Android applications. To ensure the adopted applications could cover most GUIs, we pick applications from 4 categories: travel, browser, photo and email. We record our manual interaction on these applications and format the interaction operation sequences into test cases. 
In our evaluations, we include MAPIT~\cite{mapit} and Roscript~\cite{roscript}, as the representatives of intrusive methods and non-intrusive methods. 
The results show that the existing intrusive and non-intrusive method have close success rate in replay test cases. Finally, we investigate the failure cases of replay of the two tools. We find that the ineffectiveness of MAPIT is due to 1) incorrectly located widgets and 2) failures on filtering out useless widgets. We further find the ineffectiveness of GUI matching technique largely limits the replay success rate of Roscript. The improvement opportunities are also discussed in our analysis.

\section{Background}

In this section, we introduce backgrounds that are closely related to this work.

\textbf{GUI Bugs.} Generally, GUI bugs are bugs that occur in applications and severely interfere user experience. According to previous work~\cite{glib}, GUI bugs occur mainly due to two reasons: 1) the rendering errors in runtime (software level) and 2) the incompatibility to current device (hardware level). For the runtime rendering errors, developers leverage functionality testing to detect bugs. However, the functionality testing is ineffective in detecting incompatibility bugs. Because compatibility bugs require observations on applications deployed in cross-device or cross-platform scenario, which are orthogonal to the design of functionality testing. As the graphically-rich interfaces become mainstream in current applications, developers adopt more GUI widgets than previous, leading to the rising difficulty of finding compatibility GUI bugs. 

\textbf{Functionality Testing.} Generally, functionality testing aims at traversing as much GUI scenes as possible to expose bugs. To cover more GUI scenes, an early attempt hires developers to write testing scripts to mimic human interactions on devices. However, this may lead to huge cost of human efforts. To automate the testing process and alleviate inefficiency, researchers have proposed a number of methods, which can be roughly categorized as: search-based methods~\cite{sapienz}, model-based methods~\cite{stoat}, state-based methods~\cite{timemachine} and deep learning based method~\cite{qtesting}. These methods largely expand the edge of automating functionality testing.
However, existing functionality testing techniques have one common limitation: relying on code instrumentation or available source code of applications. Usually, the source code of applications is unavailable. In such case, the functionality testing technique may be ineffective~\cite{industryGUI2022}.

\textbf{Compatibility Testing.} Compatibility testing aims at reducing bugs resulting from device or platform difference, and further keeping a uniform user experience. When the smart phones just come into being, the compatibility testing is largely overlooked because devices are similar to each other. Along with the rapid development of software and hardware, the rapidly evolved variety of smart phones~\cite{smartphoneshipment}, as well as the increased number of devices make incompatibility bugs occur more frequently than ever. This evolution poses large challenges for compatibility testing developers. For some cases, the device screen is extended. For example, devices from Samsung Galaxy S series have screen edges~\cite{samsung}. The developers must adjust applications to fit the edge screen. In other cases, the GUI widgets are buried by the front cameras, because the design of front camera varies among devices~\cite{frontcamera}. When compatibility testing encountering the above issues, the existing test cases become ineffective. Developers must rewrite new test cases to fit new devices, and this is the main reason of inefficiency.


\section{Approach Overview}

In this section, we discuss the technical details and implementations of intrusive and non-intrusive methods. Based on our inspections, we summarize and compare their difference. 

\begin{figure}[b]
    \centering
    \includegraphics[width=0.5\textwidth]{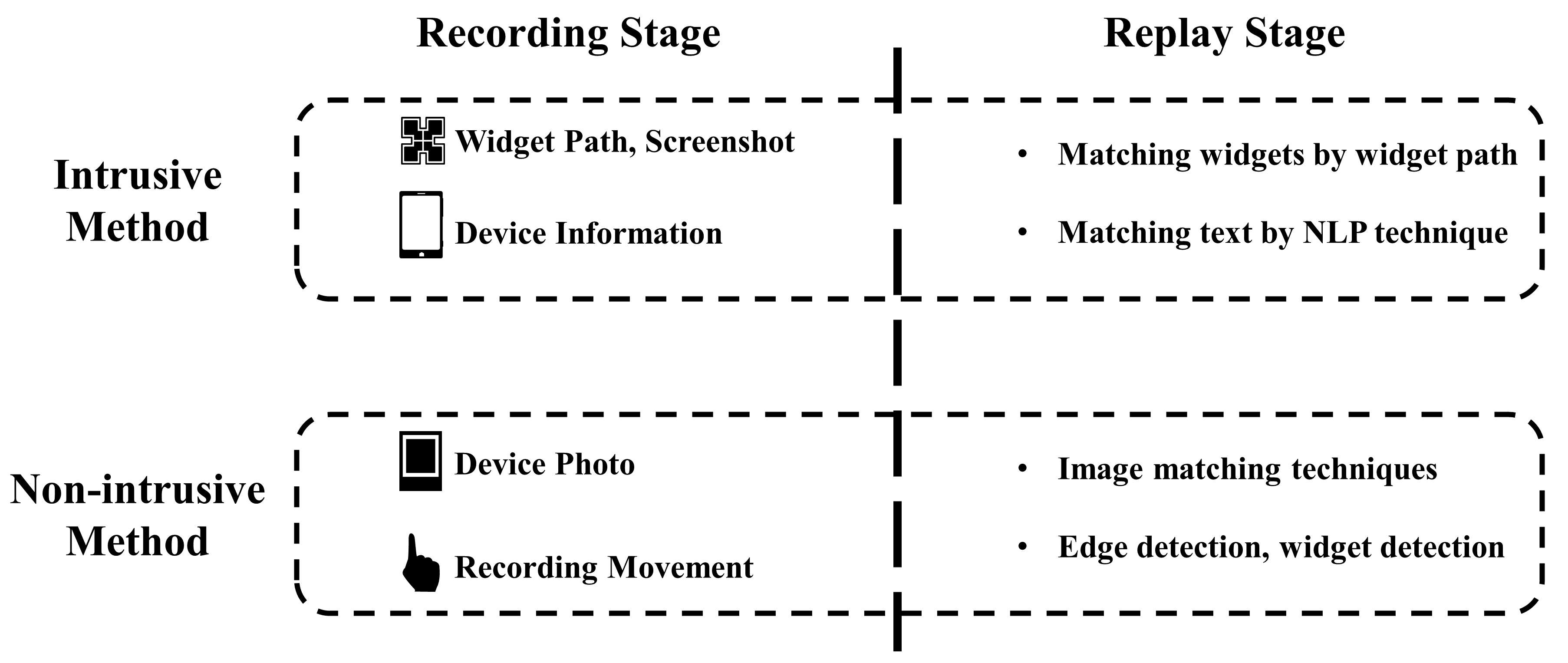}
    \caption{Comparison of intrusive method and non-intrusive method}
    \label{fig:approach compare}
\end{figure}

\subsection{Intrusive Approach}

For intrusive approach, we take MAPIT~\cite{mapit} and LIRAT~\cite{lirat}, the two state-of-the-art tools, as examples. Generally, the intrusive approach includes two stages: recording stage and replay stage. For recording stage, the tool records the test case and save the necessary information (e.g., the actions, the target widget) in particular format. And for replay stage, the tool replays the recorded test case in other devices or other platforms.

\textbf{Recording Stage.} In the implementation of intrusive methods, they first connect device by ADB (for Android) and WDA (for iOS). Once the testing developers operate on devices, the tool extract the operation coordinates as well as context information (e.g., widget path, screen size of device). As explained in LIRAT, they build an 8-tuple list for every operation, the list includes: screenshot, cropped widget screenshot, widget coordinate, device serial number, device resolution, operation coordinate, widget text and operation type. In MAPIT, though the recording steps are not well explained, we take an investigation on their implementation, and we find that the MAPIT shares similar design of LIRAT. 

\textbf{Replay Stage.} In replay stage, existing tools leverage techniques to match elements by widget information. Specifically, LIRAT adopts visual matching techniques in widget matching. This tool first extract features from widgets of target GUI to obtain candidate widgets, and then process widgets by using Canny Edge Detection Algorithm to conduct fine-grained widget characterization matching. Comparatively, MAPIT pursues matching widgets in semantic level. This tool attempts to match existing test scenarios (e.g., account registration, login) in different applications. 

\subsection{Non-intrusive Approach}

For non-intrusive approach, the state-of-the-art is Roscript~\cite{roscript}. However, after we contact the authors, we receive replies that the authors do not have plans of open-source. Therefore, we re-implement their methods ourselves, and we try to keep the implementation details the same with that introduced in their paper. 

\textbf{Recording Stage.} As shown in Figure~\ref{fig:approach compare}, unlike intrusive approach that adopts ADB to extract layout information from applications, Roscript attempts to replace this by leveraging a camera to capture visual information. Specifically, they hang a camera upon so that the device screen is placed right down the camera. When recording starts, the screen of the device is cropped, and then the developer interacts with the device to record test cases. The movements of the developer's hand are recorded and analyzed for predicting the coordinates of interacted widget. After we further investigate this method, we find the effectiveness of recording can be easily affected by environment factors (e.g., hand color, the desk color). Therefore, this method has limitations to environment adjustments.

\textbf{Replay Stage.} In replay stage, as the non-intrusive methods do not require extracting information from GUI, they usually adopt image matching techniques to visually search for similar widgets. Based on the matching results, the method obtains the coordinates of target widget. The non-intrusive methods rely heavily on the precision of GUI matching techniques. A given incorrect matching result will usually lead to a replay failure. In previous non-intrusive method, Roscript adopts template matching from OpenCV for GUI matching. 
\section{Evaluation}

In our evaluation, we collect tools and evaluate their performance of test case replay. Note that though these two methods support test case recording, we only evaluate their performance of test case replay. The reason is: The recording experience is highly related to subjective user experience, which may produce bias in experiments. 
For the baseline comparison, we include the intrusive method MAPIT~\cite{mapit} and non-intrusive method Roscript~\cite{roscript} in our evaluation. The LIRAT~\cite{lirat} is not included in our evaluation because the source code of this method is not open. Meanwhile, considering this method is similar to MAPIT, we only include MAPIT as the representative of intrusive methods. For the Roscript, though this method is unavailable, we try our best to re-implement this method following their paper. Our team consists of three developers, and we run our experiments on a computer which is running Ubuntu 18.04 LTS and equipped with Intel Xeon E5-2620v4, 32GB memories and 2TB HDD.


\begin{table}[b]
\caption{Applications adopted in our experiments. The actions on widgets include click, long-press and text input actions.}
\label{tbl: application}
\small
\centering
\begin{tabular}{cccc}
\toprule
Category & App & \# of Test Case & \# of Actions \\
\midrule
Travel & Booking & 4 & 17\\
Travel & AirBNB & 4 & 19\\
Browser & Chrome & 4 & 20 \\
Browser & Firefox & 4 & 16\\
Photo & Instagram & 4 & 17\\
Photo & Pinterest & 3 & 14\\
Email & Gmail & 4 & 16\\
Email & Outlook & 3 & 11\\
\bottomrule
\end{tabular}
\end{table}

Our experiments aim at answering the following two research questions (RQs):
\begin{enumerate}
    \item[\textbf{RQ1}] How do the tools perform in experiments? What about their success rate in test case replay?
    \item[\textbf{RQ2}] What are the challenges? Are there improvement opportunities?
\end{enumerate}


\subsection{RQ1: Evaluation of Existing Methods}

To answer RQ1, we run experiments and collect success cases and failure cases of each tool. Based on this, we calculate success rate of the tools. Additionally, we build a dataset for our experiments. This dataset adopts applications with high rates on Google Play Store. Meanwhile, the selected applications are from 4 different categories (e.g., travel application, browser) so that we can evaluate the methods on various types of applications. When recording test cases, we interact with the applications and record our interaction as a single step. Then we collect all steps as step sequences and aggregate them into test cases. Each step of the operation sequences contains three elements: application name, action type, action coordinate. The application name is for matching corresponding applications. The action type element records three types of actions: click, long-press and text input. The details of the applications can be found in Table~\ref{tbl: application}.

\begin{figure}[t]
    \centering
    \includegraphics[width=0.5\textwidth]{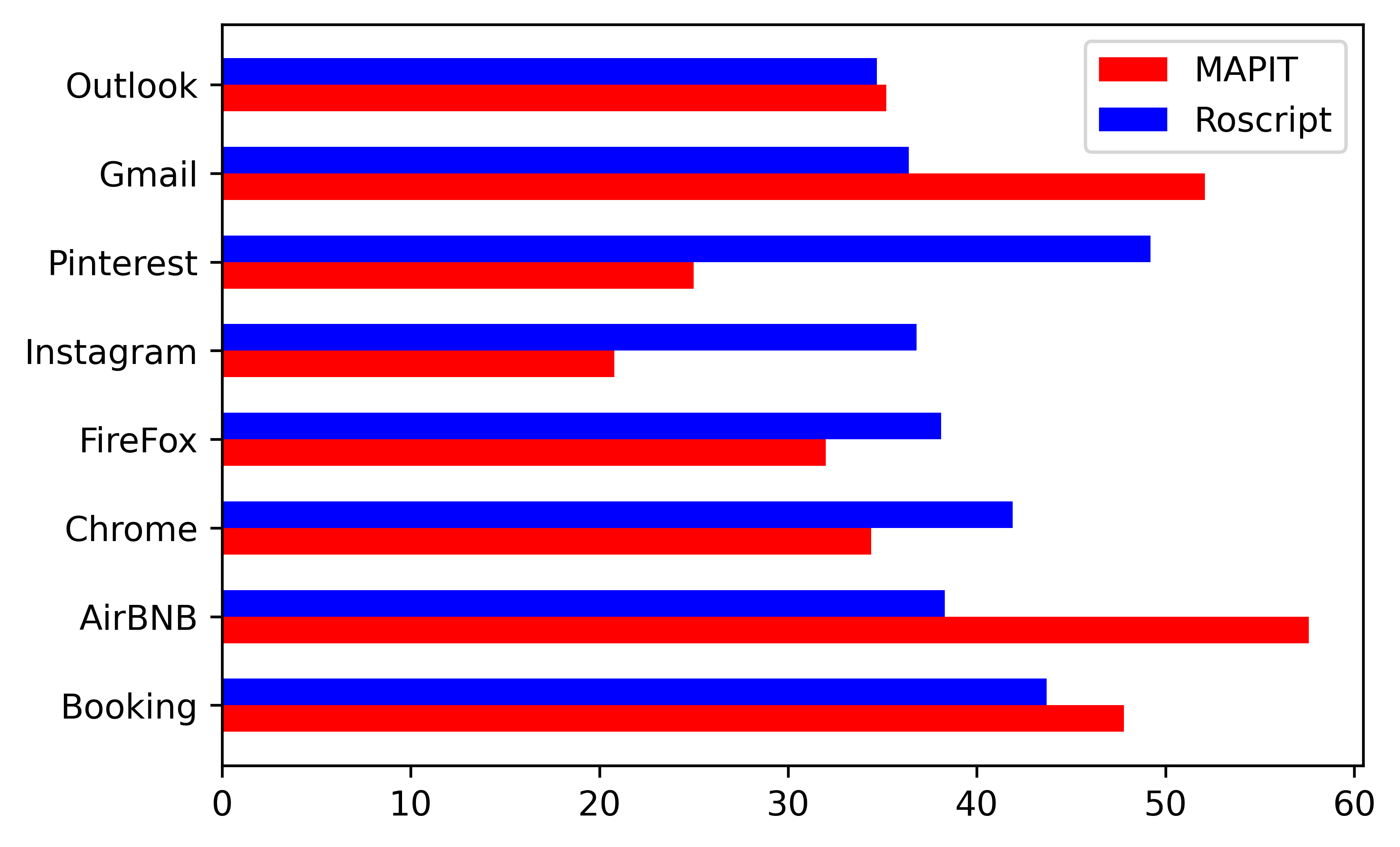}
    \caption{The performance of intrusive method and non-intrusive method on applications. The x-axis is the success rate of each step and the y-axis is the name of tested applications. The red bars stand for performance of MAPIT while the blue ones are Roscript.}
    \label{fig:eval}
\end{figure}

The results of the experiments can be found in Figure~\ref{fig:eval}. From this figure, we observe that the MAPIT outperforms Roscript on traveling applications (i.e., AirBNB and Booking) and mail applications (i.e., Outlook and Gmail). On the other hand, the Roscript outperforms MAPIT on photo applications (i.e., Pinterest and Instagram) and browser applications (i.e., FireFox and Chrome). From the perspective of overall success rate, neither of the two tools show obvious advantages to the other. Instead, the two methods have close performance. Additionally, none of the two methods can reach 60\% success rate on replaying test cases.

\subsection{RQ2: The Challenges and Opportunities}

From the experiments of RQ1, we find that the intrusive method has close replay success rate with the non-intrusive method. In this RQ, we will discuss the challenges and opportunities based on our investigations on the experiments.

\textbf{For Intrusive Method.} The failure case of MAPIT can be mainly grouped into two kinds: 1) The failure of correctly locating widgets on GUIs and 2) the difficulty in filtering out useless widgets. The summary of failure cases of MAPIT can be found in Table~\ref{tbl: cases}. The first kind of failure is that the tool MAPIT usually fails to find correctly matched widgets on target GUI. This is due to the limit of the method of MAPIT. As we investigate the code of MAPIT, we find that MAPIT locates widgets only based on the widget path and the widget description (this description is invisible for users). However, we find that there are a number of widgets that share same path and descriptions. In this case, MAPIT produces incorrect locating result and causes failure. The two elements are not enough for correctly locating a widget. Further, incapability of precisely locating a widget may be a common limit of intrusive methods.

\begin{table}[b]
\caption{MAPIT's failure cases categorization. The ``IL'' means incorrect localization, and ``FF'' represents the failure on filtering out widgets.}
\label{tbl: cases}
\small
\centering
\begin{tabular}{ccccc}
\toprule
App Name & \# of IL & \% & \# of FF & \% \\
\midrule
Outlook & 7 & 63.6\% & 4 & 36.3\% \\
Gmail & 11 & 100\$ & 0 & 0\% \\
Pinterest & 12 & 80\% & 3 & 20\% \\
Instergram & 11 & 57.8\% & 8 & 42.1\% \\
FireFox & 12 & 70.5\% & 5 & 29.4\% \\
Chrome & 11 & 57.8\% & 7 & 36.8\% \\
AirBNB & 11 & 100\% & 0 & 0\% \\
Booking & 11 & 91.6\% & 1 & 8.3\% \\
\midrule
Sum & 86 & 75.4\%  & 28 & 24.5\% \\
\bottomrule
\end{tabular}
\end{table}

\begin{figure}
    \centering
    \begin{minipage}{0.45\linewidth}
    \centering
    \includegraphics[width=\linewidth]{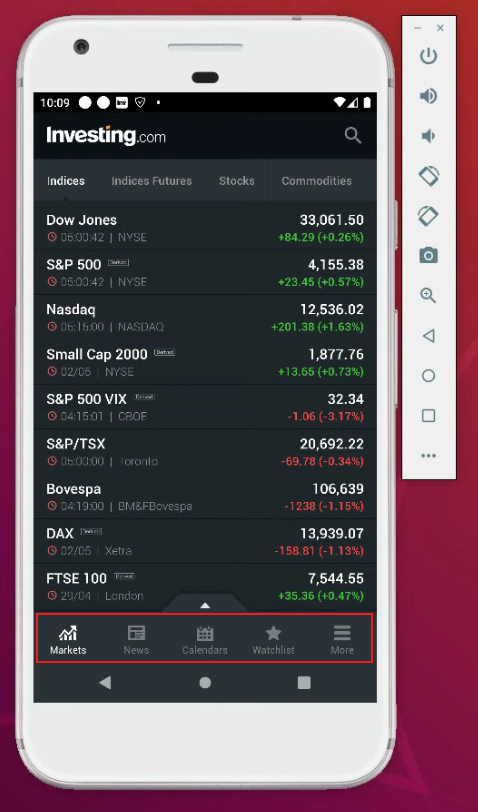}
    \caption*{(a)}
    \end{minipage}
    \begin{minipage}{0.45\linewidth}
    \centering
    \includegraphics[width=\linewidth]{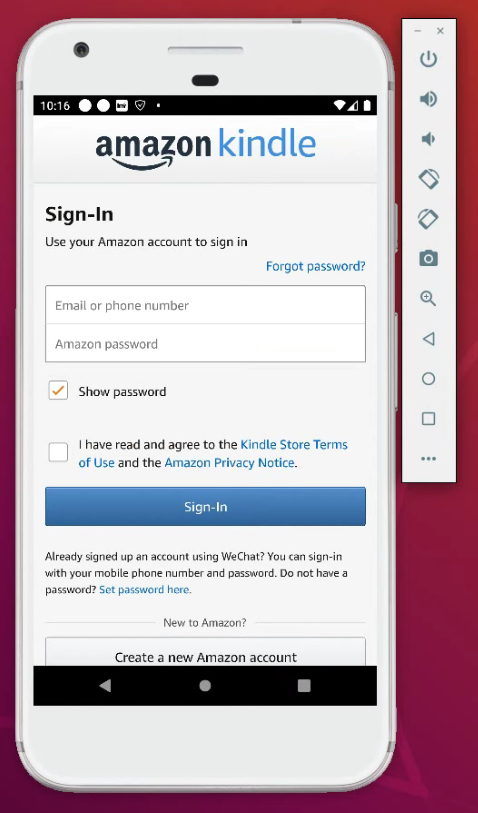}
    \caption*{(b)}
    \end{minipage}
    \caption{The two examples of overlooked widgets. In (a), the widgets in red box are all missed; in (b), the widgets in the web-view page are all overlooked.}
    \label{fig:case study}
\end{figure}

The second kind of failure is due to the difficulty in filtering out useless widgets. Generally, the efforts of analyzing all widgets to search for matching results are overwhelming. Therefore, how to minimize search space is a challenge for existing tools. MAPIT address this problem by only traversing interactable widgets instead of all widgets. In the implementation, the criterion of interactable widget is that the current widget is marked as interactable. However, in practise the application developers do not always mark an interactable widget as ``interactable''. These widgets are overlooked by MAPIT.
Further, this mistake leads to many incorrect results. The examples of overlooked widgets can be found in Figure~\ref{fig:case study}. 

\textbf{For Non-intrusive Method.} The failure cases of Roscript are mainly due to the ineffective widget matching technique adopted in replay. Specifically, Roscript uses template matching to search similar widgets in target GUI. This method often produces incorrect results. As surveyed in previous work~\cite{uied}, this old-fashioned technique show poor performance in precisely finding similar widgets and further leads to failures in test case replay. To improve this method, adopting the GUI detection technique may be a good choice.

\subsection{Threats of Validity}

We adopt two tools (i.e., MAPIT and Roscript) in our experiments. To avoid the bias introduced in the NLP prediction of MAPIT, we repeat the experiments for five times, and we calculate the average of them as the final results. Additionally, the tool Roscript is not open-sourced. We re-implement this tool by following their paper. There may be a bit of performance gap between theirs and ours. 
\section{Future Roadmap}

\textbf{Enhancing Dataset.} In current dataset, we collect 8 applications, including 30 test cases for experiments. However, this dataset is not enough for a large-scale benchmark building. Particularly, our dataset only covers 4 categories: travel, browser, photo and email applications. Other categories of applications, such as game, map, shopping, finance, are not included. Including a complete set of application categories can not only help us better find problem, but also expose limit of tools. For the dataset, our next step is to first investigate highly rated applications and expand the categories to download more applications. Additionally, we will hire more test developers to record more test cases. The test cases should cover most GUIs of the application.

\textbf{Comprehensive Evaluation.} In this work, we compare the performance of the state-of-the-art tools from non-intrusive and intrusive methods. Due to some openness issues, we only adopt MAPIT in our experiments for intrusive methods, and we re-implement Roscript as the representative of non-intrusive methods. To make our evaluations more convincing, we need to include more methods in our experiments. In our next steps, we will contact the authors and ask whether they are willing to open the source code of tools. On the other hand, if we cannot obtain the tools, we will discuss with the authors in order to ensure our re-implement tools very close to the original one.  



\newpage

\bibliographystyle{IEEEtran}
\bibliography{ref}

\end{document}